\documentclass[12pt]{iopart}

\usepackage{iopams}
 \usepackage{graphicx}
 \usepackage{epsfig}
\begin{document}

\title[Phase transition and computational complexity in a stochastic prime number generator]
{Phase transition and computational complexity in a stochastic
prime number generator}

\author{L Lacasa, B Luque}

\address{Dpto. de Matem\'atica Aplicada y Estad\'istica\\
ETSI Aeron\'auticos\\ Universidad Polit\'ecnica de Madrid.}
\ead{lucas@dmae.upm.es}

\author{O Miramontes}

\address{Dpto. de Sistemas Complejos, Instituto de F\'isica\\
 Universidad Nacional Aut\'onoma de M\'exico\\ Mexico 01415 DF, Mexico.}

\begin{abstract}
We introduce a prime number generator in the form of a stochastic
algorithm. The character of such algorithm gives rise to a
continuous phase transition which distinguishes a phase where the
algorithm is able to reduce the whole system of numbers into primes
and a phase where the system reaches a frozen state with low prime
density. In this paper we firstly pretend to give a broad
characterization of this phase transition, both in terms of
analytical and numerical analysis. Critical exponents are
calculated, and data collapse is provided. Further on we redefine
the model as a search problem, fitting it in the hallmark of
computational complexity theory. We suggest that the system belongs
to the class \emph{NP}. The computational cost is maximal around the
threshold, as common in many algorithmic phase transitions,
revealing the presence of an easy-hard-easy pattern. We finally
relate the nature of the phase transition to an average-case
classification of the problem.
\end{abstract}


\maketitle

\section{Introduction}

Computer science and physics, although different disciplines in
essence, have been closely linked since the birth of the first one.
More recently,
 computer science has met together with statistical
physics in the so called combinatorial problems and their relation
to phase transitions and computational complexity (see \cite{SFI}
for a compendium of recent works). More accurately, algorithmic
phase transitions (threshold property in the computer science
language),
 i.e. sharp changes in the behavior of some
computer algorithms, have attracted the attention of both
communities \cite{FIRST,MERT,SAT,TF,TF2,TF3,TF4,TF5}. It has been
shown that phase transitions play an important role in the resource
growing classification of random combinatorial problems \cite{TF}.
The computational complexity theory is therefore nowadays
experimenting a widespread growth, melting different ideas and
approaches coming either from theoretical computation, discrete
mathematics, and physics. For instance, there exist striking
similarities between optimization problems and the study of the
ground
states of disordered models \cite{PAR}. \\
Problems related to random combinatorics appear typically in
discrete mathematics (graph theory), computer science (search
algorithms) or physics (disordered systems). The concept of sudden
change in the behavior of some variables of the system is intimately
linked to this hallmark. For instance, Erd$\ddot{o}$s and Renyi, in
their pioneering work on graph theory \cite{erdos}, found the
existence of \emph{zero-one} laws in their study of cluster
generation. These laws have a clear interpretation in terms of phase
transitions, which appear extensively in many physical systems. More
recently, computer science community has detected this behavior in
the context of algorithmic problems. The so called \emph{threshold
phenomenon} \cite{SFI} distinguishes zones in the phase space of an
algorithm where the problem is, computationally speaking, either
tractable or intractable. It is straightforward that these three
phenomena can be understood as a unique concept, in such a way that
building bridges between each other is an
appealing idea.\\
Related to the concept of a phase transition is the task of
classifying combinatorial problems. The theory of computational
complexity distinguishes problems which are tractable, that is to
say, solvable in polynomial time by an efficient algorithm, from
those which are not. The so-called $NP$ class gathers problems
that can be solved in polynomial time by a non-deterministic
Turing machine \cite{MERT}. This class generally includes many
hard or eventually intractable problems, although this
classification is denoted \emph{worst-case}, that is to say, a
rather pessimistic one, since the situations that involve long
computations can be eventually rare. In the last years, numerical
evidences suggest the presence of the threshold phenomenon in $NP$
problems. These phase transitions may in turn characterize the
\emph{average-case} complexity of the associated problems, as
pointed out recently \cite{TF}.\\\\

In this paper we discuss a stochastic algorithm inspired on
artificial chemistry models \cite{DIT, DIT2} that has already been
studied from a statistical physics point of view \cite{Luque}.
This algorithm generates prime numbers by means of a stochastic
integer decomposition. A continuous phase transition has been
detected and described in a previous work \cite{Luque}: we can
distinguish a stationary phase where
 the ability for producing primes is practically null and a stationary phase where the algorithm is able to reduce
 the whole system into primes. It is straightforward to reinterpret the model as a search problem \cite{TF2} which
undergoes
 an algorithmic phase transition related to
 a change in its computational complexity. In this paper we firstly
 pretend to make a broader characterization of the system; in this
 sense this work is a continuation of a previous one \cite{Luque}.
 Further on, we will situate the model in the context of the
 computational complexity theory, in order to relate its computational complexity with the phase transition present in the system.
 The paper thus goes as follows:
 in section II we will
describe the model, which stands as a stochastic prime number
generator. In section III we will characterize the phase
transition present in the system, following the steps depicted in
a previous work \cite{Luque} and providing some new data and
additional information. Concretely, we will firstly outline the
breaking of symmetry responsible for the order-disorder
transition. After defining proper order and control parameters,
critical exponents will be calculated numerically from extensive
simulations and finite size scaling analysis. An analytical
approach to the system will also be considered at this point, in
terms of an annealed approximation. In section IV, we will
reinterpret the model as a search problem. We will then show that
the system belongs to the $NP$ class in a worst-case
classification. We will find an easy-hard-easy pattern in the
algorithm, as common in many $NP$ problems, related in turn to the
critical slowing down near the the transition point. According to
\cite{TF}, we will finally relate the nature of the phase
transition with the average-case complexity of the problem. We
will show that while the problem is in $NP$, the resource growing
is only polynomial. In section V we will conclude.

\section{The model}
The fundamental theorem of arithmetic \cite{primes} states that
every integer greater than one can be expressed uniquely as a
product of primes or powers of primes. In a certain manner, prime
numbers act as atoms in chemistry, both are irreducible. This way,
composite numbers can be understood as molecules. Following this
resemblance, the next algorithm has been introduced \cite{DIT}:
suppose that we got a pool of positive integers $\{2,3,...,M\}$,
from which we randomly extract a certain number $N$ of them (this
will constitute the system under study). Note that the chosen
numbers can be repeated, and that the integer $1$ is not taken into
account. Now, given two numbers $n_i$ and $n_j$ taken from the
system of $N$ numbers, the algorithm collision rules are the
following:
\begin{itemize}
\item Rule 1: if $n_i=n_j$ there is no reaction (elastic shock), and the numbers are not modified.
\item Rule 2: If the numbers are different (say $n_i > n_j$), a reaction will take place only if $n_j$ is a
divisor of $n_i$, i.e. if $n_i$ mod $n_j = 0$. The reaction will
then stand for
 \begin{equation}
 n_i \oplus n_j \longmapsto n_k \oplus n_j, \nonumber
 \end{equation}
 where $\oplus$ stands for the usual notation of a chemical reaction and $n_k = \frac{n_i}{n_j}$.
 \item Rule 3: if $n_i > n_j$ but $n_i$ mod $n_j\neq0$, no reaction takes place (elastic shock).
\end{itemize}
The result of a reaction will be the extinction of the composed
number $n_i$ and the introduction of a 'simpler' one, $n_k$.\\
The algorithm goes as follows: after randomly extracting from the
pool $\{2,3,...,M\}$ a set of $N$ numbers, we pick at random two
numbers $n_i$ and $n_j$ from the set: this is equivalent to the
random encounter of two molecules. We then apply the reaction rules.
In order to have a parallel updating, we will establish $N$
repetitions of this process ($N$ Monte Carlo steps) as a time step.
Note that the reactions tend to separate numbers into its
irreducible elements, as molecules can be separated into atoms.
Hence, this dynamic when iterated may generate prime numbers in the
system. We say that the system has reached stationarity when every
collision is elastic (no more reaction can be achieved), whether
because every number has become a prime or because rule 2 cannot be
satisfied in any case -frozen state-. The algorithm then stops.

\section{Phase transition}

\subsection{Preliminary insight}

\begin{figure}[h] 
\centering
\includegraphics[width=0.45\textwidth]{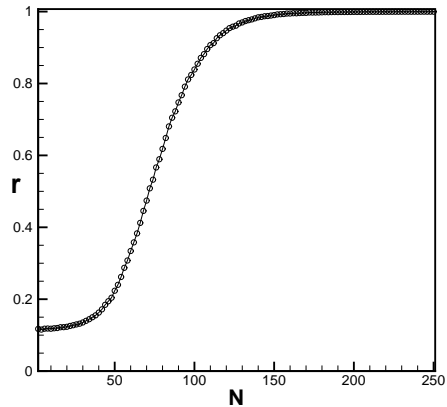}
\caption{Numerical simulation of the steady values of $r$ versus
$N$, for a pool size $M=2^{14}$. Each run is averaged over
$2\cdot10^4$ realizations in order to avoid fluctuations. Note
that the system exhibits a phase transition which distinguishes a
phase where every element of the system becomes a prime in the
steady state and a phase with low prime density.}\label{r_y_v}
\end{figure}

As stated in the previous section, this algorithm clearly tends to
generate primes as far as possible: when the algorithm stops, one
may expect the system to have a large number of primes or at least
have a frozen state of non-divisible pairs. A first indicator that
can evaluate properly this feature is the unit percentage or ratio
of primes $r$, that a given system of $N$ numbers reaches at
stationarity \cite{Luque}. In figure \ref{r_y_v} we present the
results of Monte Carlo simulations calculating, as a function of $N$
and for a concrete pool size $M=2^{14}$, the steady values of $r$.
Every simulation is averaged over $2\cdot10^4$ realizations in order
to avoid fluctuations. We can clearly distinguish in figure
\ref{r_y_v} two phases, a first one where $r$ is small and a second
one where the prime number concentration reaches the unity.
\begin{figure}[h] 
\centering
\includegraphics[width=0.75\textwidth]{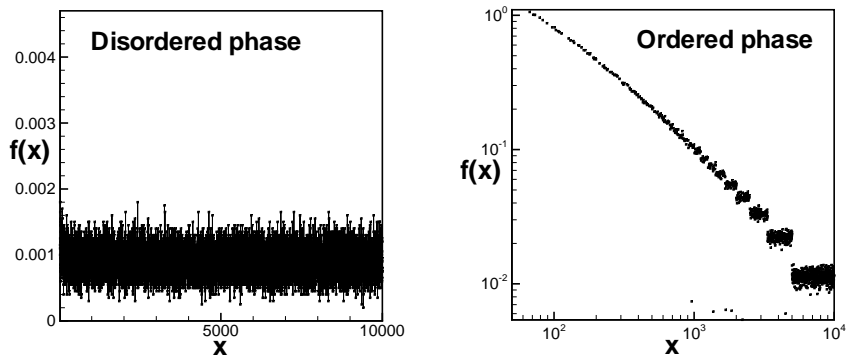}
\caption{The left figure stands for the steady state distribution
(averaged over $2\cdot 10^4$ realizations) of the $N$ elements, for
$N=10$ and $M=10^4$ (phase with low $r$): this one is a uniform
distribution U(2,M) (note that the distribution is not normalized).
The right figure stands for the same plot for $N=110$ and $M=10^4$
(phase where $r$ reaches the unity): this one is a power law
$P(x)\sim1/x$.} \label{distris}
\end{figure}
This is the portrait of a phase transition, where $N$ would stand as
the control parameter and $r$ as the order parameter. In the phase
with small $r$, the average steady state distribution of the $N$
elements is plotted in the left side of figure (\ref{distris}): the
distribution is uniform (note that the vertical scale is zoomed in
such a way that if we scale it between $[0,1]$ we would see a
horizontal line), which is related to an homogeneous state. In this
regime, every number has the same probability to appear in the
steady state. In the other hand, the average steady state
distribution of the $N$ numbers in the phase of high $r$ is plotted
in the right side of figure (\ref{distris}): the distribution is now
a power law, which is related to a biased, inhomogeneous state. In
this regime, the probability of having -in the steady state- a
composite number is practically null, and the probability of having
the prime $x$ is in turn proportional to $1/x$ \cite{comentario1}.
\begin{figure}[h] 
\centering
\includegraphics[width=0.45\textwidth]{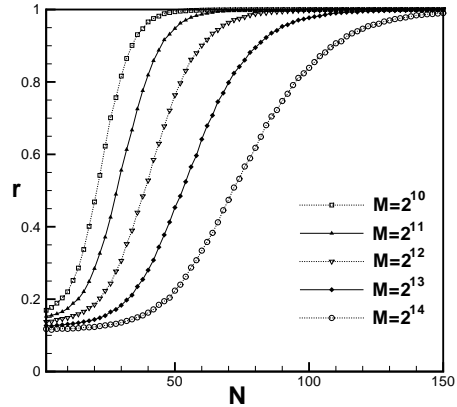}
\caption{Plot of $r$ versus $N$, for different pool sizes $M$ (each
simulation is averaged over $2\cdot10^4$ realizations).} \label{r2}
\end{figure}
The breaking of this symmetry between steady distributions leads us
to assume an order-disorder phase transition, the phase with small
proportion of primes being the disordered phase and the ordered
phase being the one where $r$ tends
to one.\\

A second feature worth investigating is the size dependence of the
transition. In figure \ref{r2} we plot $r$ versus $N$, for a set of
different pool sizes $M$. Note that the qualitative behavior is size
invariant, however the transition
point increases with $M$. This size dependence will be considered in a later section.\\
\begin{figure}[h] 
\centering
\includegraphics[width=0.45\textwidth]{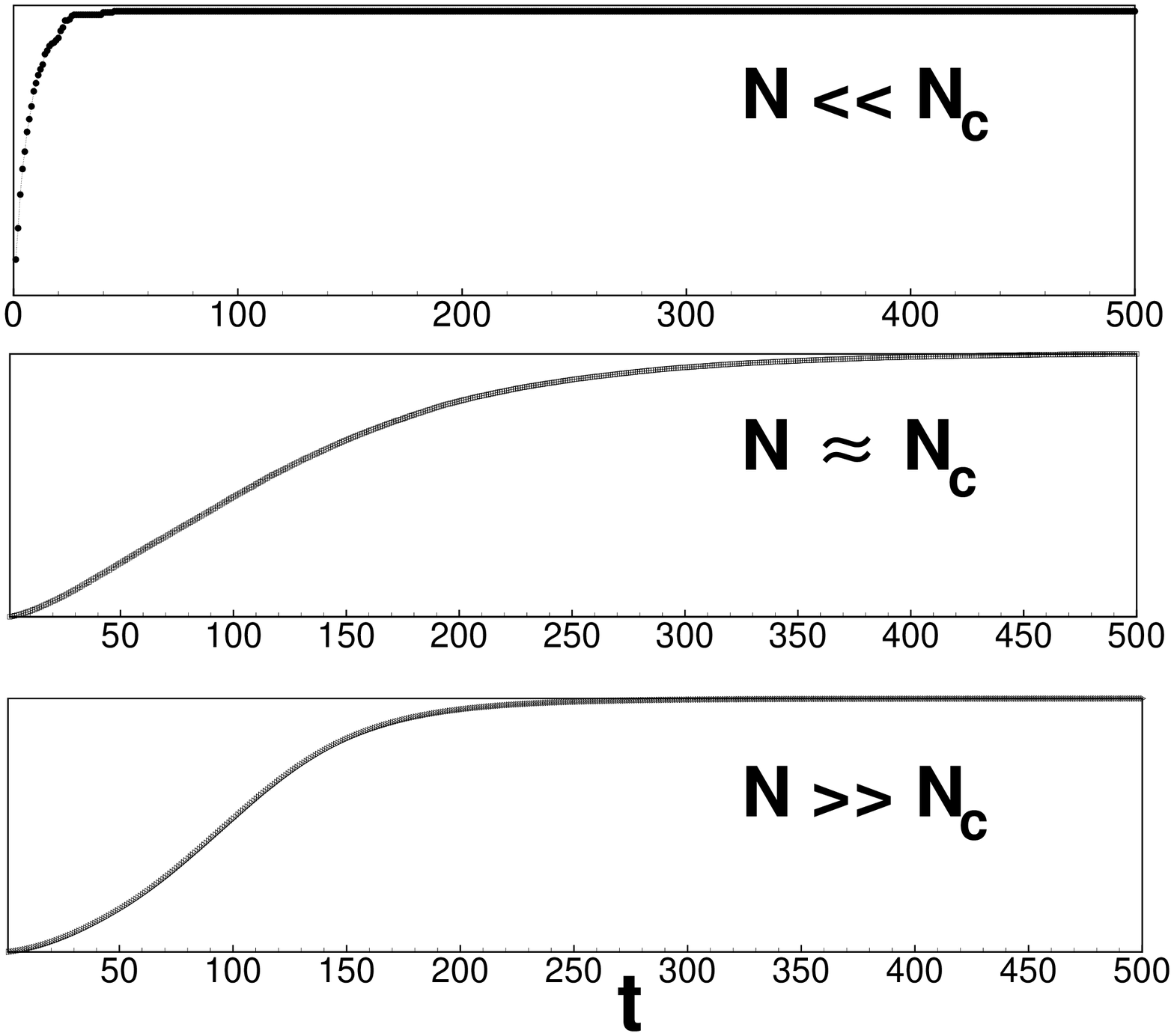}
\caption{Number of cumulated positive reactions (rule 2 is
satisfied) as a function of the time steps, for three different
configurations: (up) disordered phase $N<<N_c$, (middle) critical
phase $N\sim N_c$, (bottom) ordered phase $N>>N_c$.} \label{series}
\end{figure}
\begin{figure}[h] 
\centering
\includegraphics[width=0.45\textwidth]{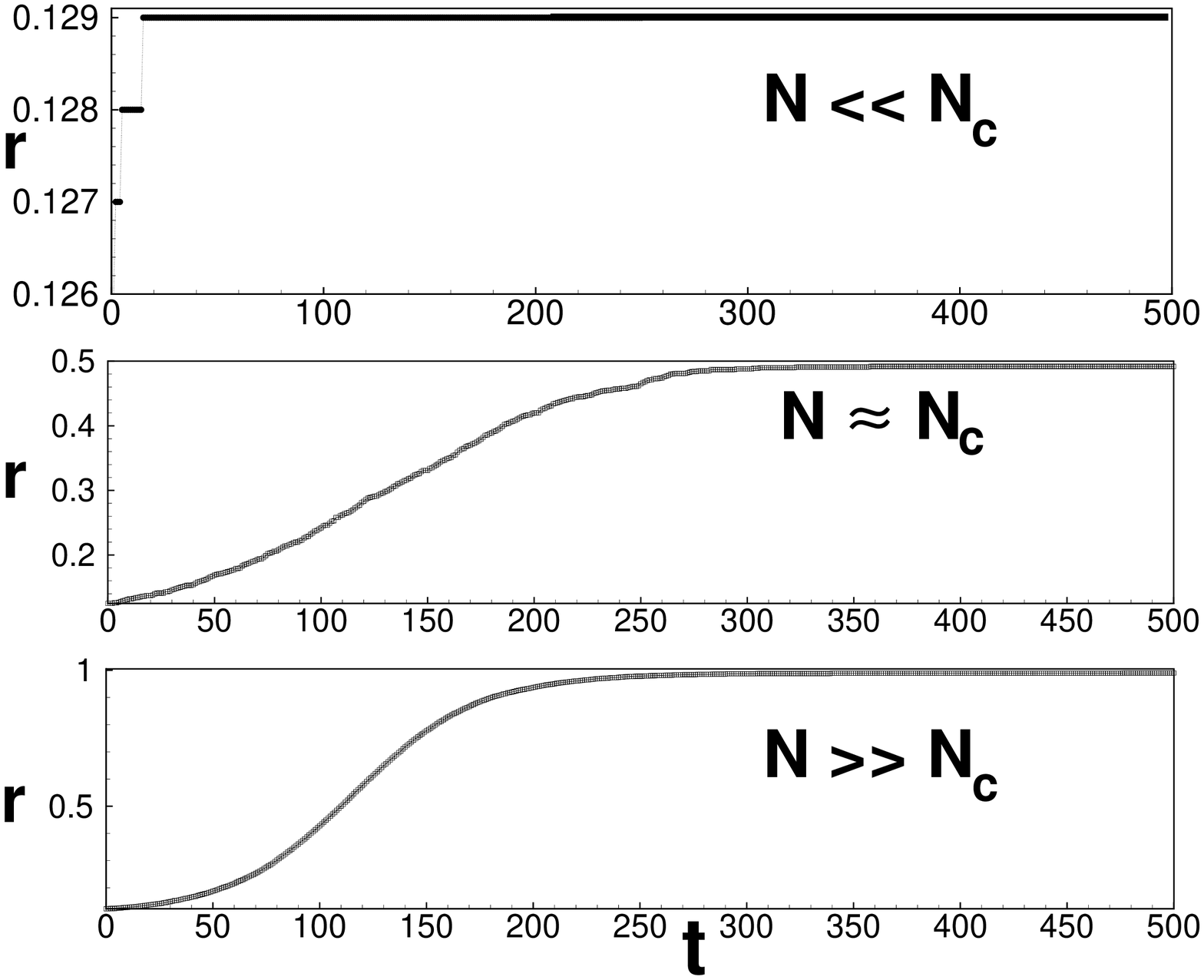}
\caption{Ratio $r$ as a function of the time steps for the same
configurations as for figure \ref{series}.} \label{concentraciones}
\end{figure}
As a third preliminary insight, we shall study the temporal
evolution of the system. In figure \ref{series} we plot, for a
given pool size $M=10^4$, the cumulated number of reactions that a
system of $N$ numbers needs to make in order to reach
stationarity. According to this, in figure \ref{concentraciones}
we plot, for the same $(N,M)$, the evolving value $r(t)$. In the
disordered phase we can see that the system is rapidly frozen: the
algorithm is not efficient in producing primes, and $r$ is
asymptotically small. In the ordered phase  the system needs more
time to reach stationarity: this is due to the fact that the
algorithm is producing many primes, as the evolving value of $r$
reaches the unity. It is however in a neighborhood of the
transition where the system takes the higher times to reach the
steady state: the system is producing many reactions, but not that
many primes. This fact can be related to a critical slowing down phenomenon, and is studied further in the text.\\

It is worth notting in figures \ref{r_y_v} and \ref{r2} that in the
disordered phase the order parameter doesn't vanish, as it should.
This is due to the fact that in a pool of $M$ numbers, following the
prime number theorem, one finds on average $M/\log(M)$ primes
\cite{primes}. Thus, there is always a residual contribution to the
ratio $1/\log(M)$ not related to the system's dynamics which only
becomes relevant for small values of $N$, when the algorithm is not
able to produce primes. This feature can be understood according to
figures (\ref{series}, \ref{concentraciones}), where in the
disordered phase the system rapidly reaches a frozen state. Note
that the ratio in this case increases from $0.126$ -residual ratio-
to an asymptotic value of $0.129$. It is clear that the system in
this phase is not able to produce primes, and that the residual
ratio
takes practically the whole contribution.\\
Since an order parameter should vanish in the disordered phase and
be non-null in the ordered phase, a new order parameter that should
avoid these problems has to be defined.

\subsection{New order parameter}

\begin{figure}[h] 
\centering
\includegraphics[width=0.45\textwidth]{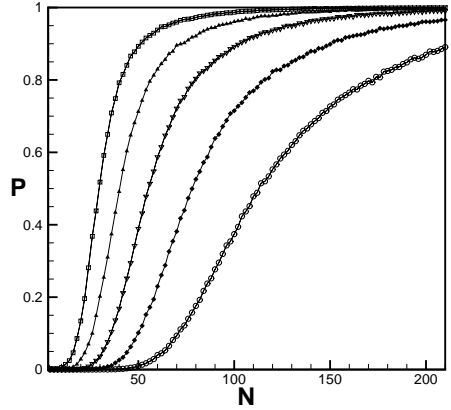}
\caption{Order parameter $P$ versus $N$, for the same pool sizes as
figure (\ref{r2}) (averaged over $2\cdot10^4$ realizations). Note
that $P$ is now a well defined order parameter, as long as $P \in
[0,1]$. Again, $N_c$ depends on the pool size $M$.} \label{P}
\end{figure}

The preliminary study suggests the presence of a phase transition
that distinguishes a phase where the algorithm is able to reduce
every element of the system into a prime from another where the
system is frozen in a state of low prime concentration. It is
straightforward that the algorithm searches stochastically the
suitable combinations through which initial numbers can react and
reduce into simpler ones.
Let us now see how this phase transition can be understood as a
dynamical process embedded in a catalytic network having integer
numbers as the nodes. Consider two numbers of that network, say
$a$ and $b$ ($a>b$). These numbers are connected ($a \rightarrow
b$) if they are exactly divisible, that is to say, if $a/b=c$ with
c being an integer. The topology of similar networks has been
studied in \cite{integers1,integers2,integers}, concretely in
\cite{integers} it is shown that this network exhibits scale-free
topology \cite{scalefree}: the degree distribution is $P(k) \sim
k^{-\lambda}$ with $\lambda=2$. In our system, fixing $N$ is
equivalent to selecting a random subset of nodes in this network.
If $a$ and $b$ are selected they may react giving $a/b=c$; in
terms of the network this means that the path between nodes $a$
and $b$ is travelled thanks to the catalytic presence of $c$. We
may say that our network is indeed a catalytic one
\cite{kauffman,kauffman2} where there are no cycles as attractors
but two different stationary phases: (i) for large values of $N$
all resulting paths sink into primes numbers, and (ii) if $N$ is
small only a few paths are travelled and no primes are reached.
Notice that in this network representation, primes are the only
nodes that have input links but no output links (by definition, a
prime number is only divisible by the unit and by itself, acting
as an absorbing node of the dynamics). When the temporal evolution
of this algorithm is explored for small values of $N$, we have
observed in figures \ref{series},\ref{concentraciones} that the
steady state is reached very fast. As a consequence, there are
only few travelled paths over the network and since $N$ is small
the probability of catalysis is small as well, hence the paths
ending in prime nodes are not travelled. We say in this case that
the system freezes in a disordered state. In contrast when $N$ is
large enough, many reactions take place and the network is
travelled at large. Under these circumstances, an arbitrary node
may be catalyzed by a large $N-1$ quantity of numbers, its
probability of reaction being high. Thus, in average all numbers
can follow network paths towards the prime nodes:
we say that the system reaches an ordered state.\\
In the light of the preceding arguments, it is meaningful to define
a new order parameter $P$ as the probability that the system has for
a given ($N$,$M$) to reduce every number from $N$ into primes, that
is to say, to reach an ordered state. In practice, $P$ is calculated
in the following way: given ($N$,$M$), for each realization we
check, once stationarity has been reached, whether the whole set of
elements are primes or not.
In the first case, $P$ sums up one, in the second, it sums up zero. We then average $P$ over every realization.\\
In figure \ref{P} we plot $P$ versus $N$, for different pool sizes
$M$. The phase transition that the system exhibits has now a clear
meaning; when $P=0$, the probability that the system has to be able
to reduce the whole system into
primes is null (disordered state), and viceversa when $P\neq0$.\\
In each case, $N_c(M)$, the critical value separating the
 phases $P=0$ and $P\neq0$, can now be defined. Observe in figure \ref{P} that $N_c$ increases with the pool
size $M$.
 In order to describe this size dependence, we need to find some analytical argument by means of which define
 a system's characteristic size. As we will see in a few lines, this one won't be
 $M$ as one would expect in a first moment.\\

\subsection{Annealed approximation}

\begin{figure}[h] 
\centering
\includegraphics[width=0.45\textwidth]{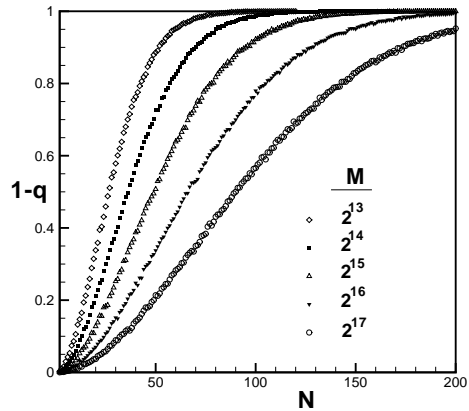}
\caption{Numerical simulations calculating the probability
$1-q(N,M)$ (as explained in the text) versus $N$, for different
values of pool size $M$, in the annealed approximation.}
\label{campo}
\end{figure}

The system under hands shows highly complex dynamics: correlations
take place between the $N$ numbers of the system at each time step
in a non trivial way. Find an analytical solution to the former
problem is thus completely out of the focus of this paper.
However, an annealed approximation can still be performed. The
main idea is to obviate these two-time correlations, assuming that
at each time step, the $N$ elements are randomly generated. This
way, we can calculate, given $N$ and $M$, the probability $q$ that
at a single time step, no pair of molecules between $N$ are
divisible. Thus, $1-q$ will be the probability that there exist at
least one reacting pair. Note that $1-q$ will somehow play the
role
of the order parameter $P$, in this oversimplified system.\\
In a first step, we can calculate the probability $p(M)$ that two
molecules randomly chosen from the pool $M$ are divisible:
\begin{equation}
p(M)=\frac{2}{(M-1)^2} \sum_{x=2}^{\lfloor
M/2\rfloor}\bigg\lfloor\frac{M-x}{x}\bigg\rfloor\approx
\frac{2\log{M}}{M},
\end{equation}
where the floor brackets stand for the integer part function.
Obviously, $1-p(M)$ is the probability that two molecules randomly
chosen are not divisible in any case. Now, in a system composed by
$N$ molecules, we can make $N(N-1)/2$ distinct pairs. However,
these pairs are not independent in the present case, so that
probability $q(N,M)$ isn't simply $(1-p(M))^{N(N-1)/2}$.
Correlations between pairs must be somehow taken into account. At
this point, we can make the following ansatz:
\begin{equation}
q(N,M)\approx \left(1-\frac{2\log{M}}{M}\right)^{N^{1/\alpha}}
\label{ansatz}
\end{equation}
where $\alpha$ characterizes the degree of independence of the
pairs. The relation $1-q(N,M)$ versus $N$ is plotted in figure
\ref{campo} for different values of the pool size $M$. Note that for
a given $M$, the behavior of $1-q(N,M)$ is qualitatively similar to
$P$, the order parameter in the real
system.\\

\begin{figure}[h] 
\centering
\includegraphics[width=0.45\textwidth]{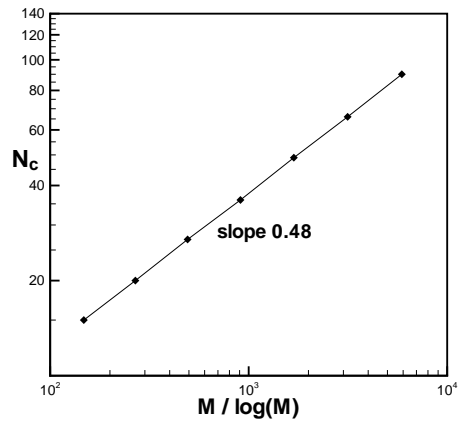}
\caption{Scaling of $N_c$ versus the system's characteristic size in
the annealed approximation. The plot is log-log: the slope of the
straight line provides the exponent $\alpha=0.48$ of equation
(\ref{escala1}).} \label{scaling1}
\end{figure}

For convenience, in this annealed approximation we will define a
threshold $N_c$ as the one for which $q(N_c,M)=0.5$. This value is
the one for which half of the configurations reach an ordered state.
This procedure is usual for instance in percolation processes,
 since the choice of the percolation threshold, related to the definition of a spanning
cluster, is somewhat arbitrary in finite size
 systems \cite{percolation}. Taking logarithms in
equation (\ref{ansatz}) and expanding up to first order, we easily
find an scaling relation between $N_c$ and $M$, that reads
\begin{equation}
N_c\sim\left[\frac{M}{\log{M}}\right]^{\alpha} .\label{escala1}
\end{equation}
 This relation suggests that the system's characteristic size is not $M$, as one would expect in a first moment,
 but $M/\log(M)$.
In figure \ref{scaling1} we plot, in log-log, the scaling between
$N_c$ and the characteristic size $M/\log(M)$ that can be
extracted from figure \ref{campo}. The best fitting provides a
relation of the shape (\ref{escala1}) where $\alpha=0.48\pm0.01$
(note that the scaling is quite good, what gives consistency to
the leading order approximations assumed in equation
\ref{escala1}).

\subsection{Data collapse and critical exponents}

\begin{figure}[h] 
\centering
\includegraphics[width=0.45\textwidth]{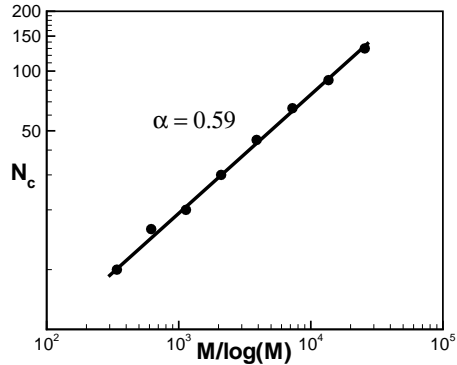}
\caption{Scaling of the critical point $N_c$ versus the
characteristic system's size $M/\log(M)$ in the prime number
generator, for pool size $M=\{2^{10}-2^{18}\}$. The plot is log-log:
the slope of the curve provides an exponent $\alpha=0.59$.}
\label{scaling2}
\end{figure}


The annealed approximation introduced in the preceding section
suggests that the characteristic size of the system is not $M$ as
one would expect but rather $M/\log(M)$. This is quite reasonable if
we have in mind that the number of primes that a pool of $M$
integers has is on average
 $M/\log(M)$ \cite{primes}: the quantity of primes doesn't grow linearly with $M$. This is the so called prime
 number theorem, and states that the
number of primes in the set of integers $\{1,2..M\}$ is asymptotic with $M/\log(M)$ when $M$ is large enough.\\
In order to test if this conjecture also applies to the prime
number generator, in figure \ref{scaling2} we represent (in
log-log) the values of $N_c$ (obtained numerically from the values
where $P(N,M)$ becomes non-null for the first time) as a function
of $M/\log(M)$. We find the same scaling relation as for the
annealed system (equation \ref{escala1}), but with a different
value for $\alpha=0.59\pm0.05$. This little disagreement is
logical and comes from the fact that in the annealed approximation
correlations are
obviated.\\
Let us apply generic techniques of finite size scaling in order to
calculate the critical exponents of this phase transition. Reducing
the control parameter as $n=\frac{N}{M/\log(M)}$, the correlation
exponent $\nu$ is defined as:
\begin{equation}
|n-n_c| \sim \bigg(\frac{M}{\log(M)}\bigg)^{-1/\nu},
\label{correlacion}
\end{equation}
where we find $n_c(\infty)=0$ and $\nu=1.69\pm0.05$. Note that the
transition tends to zero in the thermodynamical limit because its
value increases more slowly than the system's size.
\begin{figure}[h] 
\centering
\includegraphics[width=0.35\textwidth]{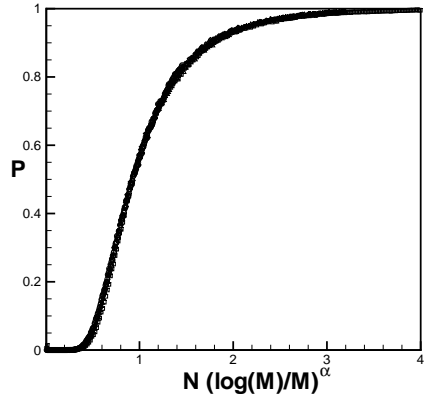}
\caption{Data collapse of curves ($N,P$) for different values of
$M$, assuming the scaling relation \ref{escala1}. The collapse is
very good, the scaling relation seems to be consistent. }
\label{colapso}
\end{figure}

The critical exponent of the order parameter $\beta$ can be
deduced from the calculation of the correlation exponent $\nu$ and
from the finite size scaling relation
\begin{equation}
P(n_c)\sim \bigg(\frac{M}{\log(M)}\bigg)^{-\beta/\nu}.
\end{equation}
The best fitting provides a value of $\beta=3.4\pm0.2$.\\
In figure \ref{colapso} we have collapsed all curves $P(N,M)$
according to the preceding developments. Note that the collapse is
excellent, something which provides consistency to the full
development.

\section{Computational complexity}

Hitherto, we have seen that the dynamical process that the prime
number generator shows gives rise to a continuous phase transition
embedded in a direct catalytic network. As pointed out in
\cite{TF2}, phase transitions quite similar to the former one as
percolation processes for instance can be easily related to search
problems. In the case under study we can redefine the percolation
process as a decision problem in the following terms: one could ask
when does the clause \emph{every number of the system is prime when
the algorithm reaches stationarity} is satisfied. It is clear that
through this focus, the prime number generator can be understood as
a SAT-like problem, as long as there is an evident parallelism
between the satisfiability of the preceding clause and our order
parameter P. Thereby, in order to study the system from the focus of
computational complexity theory, we must tackle the following
questions: what is the algorithmic complexity of the system? and how
is related the observed phase transition to the problem's
tractability?

\subsection{Worst-case classification}
The algorithm under study, which generates primes by stochastic
decomposition of integers, is related to both primality test and
integer decomposition problems. Although primality was believed to
belong to the so-called $NP$ problems \cite{PR} (solvable in
non-deterministic polynomial time), it has recently been shown to be
in $P$ \cite{primos2}: there exists at least an efficient
deterministic algorithm that tests if a number is prime in
polynomial time. The integer decomposition problem is in turn a
harder problem, and to find an algorithm that would factorize
numbers in polynomial time is an unsolved problem of computer
science. Furthermore, exploring the computational complexity of the
problem under hands could
eventually shed light into these aspects.\\
For that task, let us determine how does the search space grows when
we increase $N$. In a given time step, the search space corresponds
to the set of configurations that must be checked in order to solve
the decision problem: this is nothing but the number of different
pairs that can be formed using $N$ numbers. Applying basic
combinatorics, the set of different configurations $G$ for $N$
elements and $N/2$ pairs is:
\begin{equation}
G(N)=\frac{N!}{2!^{N/2}.(N/2)!}=(N-1)!!.
\end{equation}
We get that the search space increases with $N$ as $(N-1)!!$. On the
other hand, note that the decision problem is rapidly checked (in
polynomial time) if we provide a candidate set of $N$ numbers to the
algorithm. These two features lead us to assume
that the problem under hands belongs, in a worst-case classification \cite{SFI}, to the $NP$ complexity class.\\
Note that this is not surprising: the preceding sections led us to
the conclusion that the process is embedded in a (dynamical)
scale-free catalytic network. As a matter of fact, the phase
transition is related to a dynamical process embedded in a high
dimensional catalytic network. In this hallmark, it is
straightforward that this underlying network is non-planar
\cite{graph}. Now, it has been shown that non-planarity in this
kind of problems usually leaves to NP-completeness \cite{ISING}
(for instance, the Ising model in two-dimensions is, when the
underlying network topology is non-planar, in $NP$).

\subsection{Easy-hard-easy pattern}

\begin{figure}[h]
\centering
\includegraphics[width=0.45\textwidth]{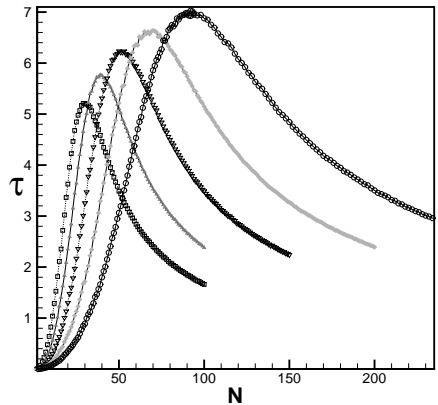}
\caption{Characteristic time $\tau$ as defined in the text versus
$N$, for different pool sizes, from left to right:
$M=2^{10},2^{11},2^{12},2^{13},2^{14}$. Every simulation is averaged
over $2\cdot10^4$ realizations. Note that for each curve and within
the finite size effects $\tau(N)$ reaches a maximum in a
neighborhood of its transition point (this can be easily explored in
figure \ref{P}).} \label{t_sin_colapso}
\end{figure}

\begin{figure}[h]
\centering
\includegraphics[width=0.45\textwidth]{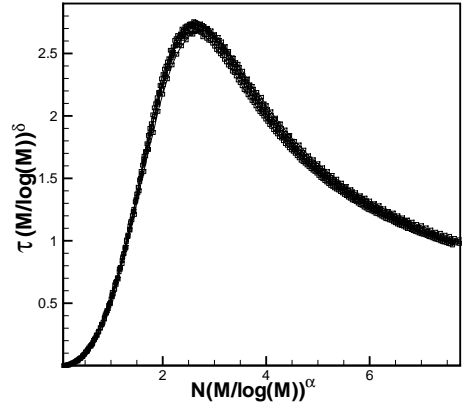}
\caption{Data collapse of $\tau$ for the curves of figure
\ref{t_sin_colapso}. The goodness of the collapse validates the
scaling relations.} \label{colapse_t}
\end{figure}

An ingredient which is quite universal in the algorithmic phase
transitions is the so called \emph{easy-hard-easy pattern}
\cite{SFI}: in both phases, the computational cost of the algorithm
(the time that the algorithm requires to find a solution, that is,
to reach stationarity) is relatively small. However, in a
neighborhood of the transition, this computational time reaches a
peaked maximum. In terms of search or decision problems, this fact
has a clear interpretation: the problem is relatively easy to solve
as long as the input is clearly in one phase or the other, but not
in between. In the system under study, the algorithm is relatively
fast in reaching an absorbing state of low concentration of primes
for small $N$ because the probability of having reactions is small.
In the other hand, the algorithm is also fast in reaching an
absorbing state of high concentration of primes for high $N$,
because the system has enough "catalytic candidates" at each time
step to be able to reduce them, the probability of having reactions
is high. In the transition's vicinity, the system is critical.
Reactions can be achieved, however, the system needs to make an
exhaustive search of the configuration space in order to find these
reactions: the algorithm requires in this region
much more time to reach stationarity.\\
Note that this easy-hard-easy pattern is related, in second order
phase transitions, to the the phenomenon of critical slowing down,
where the relaxation time in the critical region
diverges \cite{SFI}.\\

 We have already seen in figure \ref{series} that the system reaches the steady state in a different manner,
 depending on which phase is located the process. More properly, when $N<<N_c$ (disordered phase), the system rapidly frozens,
 without practically achieving any reaction. When $N>>N_c$ (ordered phase), the system takes more time to reach the steady state,
 but it is in the regime $N\sim N_c$ where this time is maximal. In order to be able to properly compare these
 three regimes, let us define a characteristic time in the system $\tau$ as the number
of average time steps that the algorithm needs to take in order to
reach stationarity. Remember that we defined a time step $t$ as $N$
Monte Carlo steps ($N$ operations). Thus, in order to normalize over
the number of molecules, it is straightforward to define a
characteristic time as:
\begin{equation}
\tau(N)=\frac{t}{N}.
\end{equation}
Note that $\tau$ can be understood as a measure of the algorithm's
time complexity \cite{MERT}. In figure \ref{t_sin_colapso} we plot
$\tau$ versus $N$ for a set of different pools $M=2^{10}...2^{14}$
(simulations are averaged over $2\cdot10^4$ realizations). Note
that given a pool size $M$, $\tau$ reaches a maximum in a
neighborhood of its transition point $N_c(M)$, as can be checked
according to figure \ref{P}. As expected, the system exhibits an
easy-hard-easy pattern, as long as the characteristic time $\tau$
required by the algorithm to solve the problem has a clear maximum
in a neighborhood of the phase transition. Moreover, the location
of the maximum shifts with the system's size according to the
critical point scaling found in equation \ref{escala1}. In the
other hand, this maximum also scales as:
\begin{equation}
\tau_{max}\bigg(M/\log(M)\bigg)\sim\bigg(M/\log(M)\bigg)^\delta,
\end{equation}
where the best fitting provides $\delta=0.13\pm0.1$. Note that in
the thermodynamic limit, the characteristic time would diverge in
the neighborhood of the transition. It is straightforward to relate
this parameter with the relaxation time of a physical phase
transition. According to these relations, we can collapse the curves
 $\tau(N,M)$ of figure \ref{t_sin_colapso} into a single universal one. In figure \ref{colapse_t} this
collapse is provided: the goodness of the former one supports the
validity of the scaling relations.

\subsection{Average-case classification}
The system under study is interpreted in terms of a search problem,
belonging to the $NP$ class in a worst-case classification. Now, an
average-case behavior, which is likely to be more useful in order to
classify combinatorial problems, turns out to be tough to describe.
In \cite{TF}, Monasson et al. showed that there where $NP$ problems
exhibit phase transitions (related to dramatic changes in the
computational hardness of the problem), the order of the phase
transition is in turn related to the average-case complexity of the
problem. More specifically,
 that second order phase transitions are related to a polynomial growing
 of the resource requirements, instead of exponential growing, associated
 to first order phase transitions. \\
It has been shown that the system actually exhibits a second order
phase transition and an easy-hard-easy pattern. Following Monasson
et al. \cite{TF}, while our prime generator is likely to belong to
the $NP$ class, it shows however only a polynomial growing in the
resource requirements, in the average case. One may argue that one
of the reasons of this hardness reduction is that the algorithm
doesn't realize a direct search but on the contrary this search is
stochastic: the search space is not exhaustively explored. Thereby,
the average behavior of the system and thus the average decision
problem can be easily solved by the algorithm, in detriment of the
probable
character of this solution.\\

\section{Conclusions}
In this paper a (stochastic) algorithmic model which stands for a
prime number generator has been studied. This model exhibits a
phase transition which distinguishes a phase where the algorithm
has the ability to reduce every element into a prime, and a phase
where the system is rapidly frozen. Analytical and numerical
evidences suggest that the transition is continuous. On a second
part, the model has been reinterpreted as a search problem. As
long as the model searches paths to reduce integers into primes,
the combinatorial problem is related to primality test and
decomposition problem. It has been shown that this model belongs
to the $NP$ class in a worst-case classification, moreover, an
easy-hard-easy pattern has been found, as common in many
algorithmic phase transitions. According to the fact that the
transition is continuous, and based on previous works, it has been
put into relevance that the average-case complexity may be only
polynomial. This hardness reduction is in turn related to the fact
that the algorithm only yields probable states.
\section{Acknowledgments}
The authors wish to thank the Instituto de Fisica at UNAM for
support and hospitality. The research was supported by UNAM-DGAPA
grant IN-118306 (OM), grants FIS2006-26382-E and FIS2006-08607 (BL
and LL) and the \emph{Tomas Brody Lectureship 2007} awarded to BL.


\begin{thebibliography}{30}
\bibitem{comentario1} This is something quite
intuitive if we take into account that there are typically $N/2$
multiples of two, $N/3$ multiples of three, and so on: the
probability that the prime $x$ appears in a random composite number
is in average $1/x$.
\end{thebibliography}
\end{document}